\documentclass[11pt]{article}
\usepackage{mwe}
\usepackage{wrapfig}
\usepackage{amssymb}
\usepackage{xcolor}
\usepackage{mathtools}
\usepackage{soul,xcolor}
\usepackage{ulem}
\usepackage{graphicx}
\usepackage{authblk}
\usepackage{cite}
\graphicspath{{./images/}}
\usepackage{hyperref}
\usepackage[utf8]{inputenc}
\usepackage{tabularx}
\usepackage{cite}

\newcommand{\ind}{$\mbox{ }$\hspace{2mm}}

\newcommand{\dind}{\ind \ind}
\newcommand{\indd}{\ind \,  \,}

\newtheorem{definition}{Definition}
\newtheorem{proposition}{Proposition}

\newtheorem{corollary}{Corollary}
\newtheorem{theorem}{Theorem}
\newtheorem{lemma}{Lemma}

\newcommand{\bX}{{\bf X}}
\newcommand{\bY}{{\bf Y}}
\newcommand{\bS}{{\bf S}}
\newcommand{\bZ}{{\bf Z}}

\newcommand{\mskt}{MS$k$T}
\newcommand{\msktf}{MS$k$T$(f)$}
\newcommand{\hmsktf}{H-MS$k$T$(f)$}
\newcommand{\hs}{\hspace{1mm}}
\newcommand{\kclq}{$(k+1)$-clique}
\newcommand{\kq}{\kclq}
\newcommand{\epf}{\hfill{$\square$}}
\newcommand{\kqq}{\kclq\hs}
\newcommand{\td}{tree-decomposition\hs}
\newcommand{\dl}{\Delta}
\newcommand{\cali}{{\cal I}}

\begin{document}

\title{Polynomial-time derivation \\
 of optimal $k$-tree topology from Markov networks}

\author[1]{\sc Fereshteh R. Dastjerdi}
\author[1]{\sc Liming Cai\footnote{To whom all correspondence should be addressed: {\tt liming@uga.edu}}}

\affil[1]{\vspace{2mm}School of Computing, University of Georgia, Athens, GA 30602}

\date{\today}
\maketitle
\begin{abstract}
Characterization of joint probability distribution for large networks of random variables remains a challenging task in data science. Probabilistic graph approximation with simple topologies has practically been resorted to; typically the tree topology makes joint probability computation much simpler and can be effective for statistical inference on insufficient data. However, to characterize network components where multiple variables cooperate closely to influence others, model topologies beyond a tree are needed, which unfortunately are infeasible to acquire. In particular, our previous work has related optimal approximation of Markov networks of treewidth $k\geq 2$ closely to the 
graph-theoretic problem of finding maximum 
spanning $k$-tree (MS$k$T), which is a provably intractable task.

This paper investigates optimal approximation of Markov networks with $k$-tree topology that retains some designated underlying subgraph. Such a subgraph may encode certain background information that arises in scientific applications, for example, about a known significant pathway in gene networks or the indispensable backbone connectivity in the residue interaction graphs for a biomolecule 3D structure. In particular, it is proved that the $\beta$-retaining MS$k$T problem, for a number of classes $\beta$ of graphs, admit $O(n^{k+1})$-time algorithms for fixed $k\geq 1$. These $\beta$-retaining MS$k$T algorithms offer efficient solutions for approximation of Markov networks with $k$-tree topology in the situation where certain persistent information needs to be retained.
\end{abstract}

\noindent

{\bf Keywords}: Markov network, joint probability, KL divergence, mutual information, $k$-tree, tree-width, $\beta$-retaining spanning graph, dynamic programming
\section{Introduction}
 To accurately model complex real world systems that exhibit relationships among random variables, it is often crucial to compute the joint probability distribution  of a set of random variables \cite{KschischangEtAl2001,BleiEtAl2017}.
  The joint distribution function $P(\bX)$, for set $\bX$ of random variables,  serves as a fundamental basis for a wide range of statistical and probabilistic modeling techniques employed in many fields, such as communication, pattern recognition, machine learning, information retrieval, databases, and learning systems \cite{Murphy2012,FosterAndKesselman2003,GionisEtAl1999}. Distribution $P(\bX)$ captures the relationships and dependencies between different variables or events within the system. It provides insights into the likelihood of specific outcomes, helps identify patterns and correlations, and enables prediction of unseen data \cite{mohseni2014rolling,MELUCCI201991}. It can be challenging, and sometimes even impossible within a reasonable time frame, to determine a single probability distribution due to the complex relationships between random variables. For example, to fully specify the $n^{th}$-order  probability distribution $P(\bX)$ of an event space defined by $n$ binary random variables, the knowledge of $2^n$ probability values would be required. Indeed, the exponential size of the event space makes  the exact learning of $P(\bX)$ to be an intractable task \cite{7869051,MELUCCI201991,chitsaz2009birna}. The challenge becomes even greater when only limited number of data samples are at hand \cite{chow1968approximating}.\par
 \par

 Various approaches to approximate $P(\bX)$ have proposed for situations where the information of random variables is given with a limited number of samples \cite{kovacs2009approximation,chow1968approximating,gene,mohseni2014rolling,montiel2013approximating,Speech,MELUCCI201991,1202220}.
 One common approach is to approximate the $n^{th}$-order $P(\bX)$  using a lower-order probability function. Lower-order approximations encompass the $m^{th}$-order
 approximations  for $m \ll n$. These approximations often involve simplifying assumptions, such as statistical independence or other constraints imposed on dependency relationships between the random variables.  By employing these techniques, it becomes possible to compute a specific joint probability distribution that aligns with the available data. Their effectiveness is contingent upon the accuracy of approximations in capturing the true distribution  $P(\bX)$. In a study by King {\it et al.} \cite{king2009mist}, the authors specifically focused on a second-order approximation ($m=2$) to approximate a multivariate probabilistic model.  The second-order approximation
considers only the pairwise dependency between variables,
reducing the computational complexity of calculating $P(X)$ to $O(n^2)$.\par

\par

The seminal work of Chow and Liu \cite{chow1968approximating} has made a significant contribution to the field of approximating $P(\bX)$ using a second-order approximation method with a tree-topology known as the Chow-Liu tree. Technically, this approximation method constructs a maximum spanning tree of the network whose topology guarantees to minimize information loss in approximating $P(\bX)$, as measured by the Kullback-Leibler divergence \cite{KullbackAndLeibler1951}. The Chow-Liu method has been proven both practical and computationally efficient, even with limited data availability  \cite{9719784,1202220}. In various applications, in the context of either physical or artificial systems, such a tree structure often serves as an effective model for approximating the underlying system\cite{tan2010learning}. As a result, the Chow-Liu tree is widely employed as a flexible and interpretable model across different tasks, including
dimensional reduction \cite{zhang2019tree}
systems biology \cite{jiao2016beyond}, 
gene differential analysis \cite{gene},
discrete speech recognition \cite{Speech},
data classification and clustering tasks \cite{clustering}, 
and time-series data analyses
\cite{STEIMER2015520}.\par

 \par

However, an approximation of probability distribution $P(\bX)$ with tree topology may be below the desired or expected level of performance when dealing with distributions of high-dimensional data.
In many real world scenarios, dependencies and relationships between variables are often more complex than what a tree structure can capture \cite{wainwright2008graphical,1202220,karger2001learning}. 
Specifically, the tree topology fails to deal with the scenario where one variable is an effect caused by collaborative work of two or more other variables  \cite{bresler2015efficiently}. Even in real world applications that exhibit tree-like (sparse) network structures, \cite{singh2015finding}, there exist critical components that may not be a tree structure. One such example is a bottleneck component of several genes working together to influence a couple of downstream genes in a cancerous pathway within a gene-network  \cite{gunduz2004cell}. Networks containing such locally coupled components while demonstrating global sparsity are common, e.g., a social network containing small groups of highly related people, an article citation network for advancement of specific research topic,  and a biological network including a cluster of interacting  proteins \cite{humphries2008network,newman}. 
\par

Hence, approximation of probability distribution $P(\bX)$ with sparse, non-tree graph topology is desirable; yet measuring the quality of approximation requires an accurate definition of network sparsity. 
Sparsity of a graph is often  loosely defined by the ratio of the number of edges to the number of vertices in the graph. Typically a graph $G=(V, E)$ is sparse if $\frac{|E|}{|V|} =O(1)$, i.e., the number of edges is at most linearly related to the number of vertices. This metric, while intuitive, may not account differences in global topology among such graphs. For example,   ``tree-like'' graphs bear very different global connectivity from those ``grid-like'' graphs, in spite of both having a linear number of edges.  The global connectivity of graphs can have an immediate impact on the tractability of graph optimization problems, which can be well quantified with the notion of graph ``tree width''. Tree-width measures how much a graph is tree-like and it is intimately related to the notion of $k$-tree   \cite{Patil1986,RobertsonAndSeymour1983,BODLAENDER19981}. Indeed, under various settings, investigations have been carried out on approximation of $P(\bX)$ with topology of graphs of small tree-width \cite{karger2001learning,Srebro2003,SzantaiAndKovacs2012}. In particular, our previous work connects the optimal approximation of $P(\bX)$ with graph topology of tree-width $k$  to the task of finding 
a maximum spanning $k$-tree (MS$k$T) from the underlying network graph \cite{chang2022optimal}. The latter task, unfortunately, is computationally intractable \cite{Leizhen1993} for all fixed $k\geq 2$.  

\par
In this paper, we investigate efficient algorithms for optimal Markov network approximation with graph topology of tree-width bounded by $k$ under natural condition arising from applications, specifically
when the desired topology of tree-width $k$ retains certain essential subnet in the underlying Markov network. We coin this the 
problem $\beta$-retaining \mskt, which finds a maximum spanning $k$-tree that retains a  subgraph (of class $\beta$) designated in the input graph. We prove that for class $\beta$ of bounded degree spanning trees, the $\beta$-retaining MS$k$T problem can be solved in polynomial-time $O(n^{k+1})$ on graphs of $n$ vertices for every fixed $k\geq 1$. We also show that the achieved time complexity upper is likely the optimal for problem $\beta$-retaining MS$k$T. In particular, we are able to reduce the classical graph problem $k$-Clique to 
$\beta$-retaining \mskt\ with a transformation that literally preserves the parameter $k$ value. Problem $k$-Clique is notoriously difficult to admit algorithms of time $O(n^{k-\epsilon})$, for $\epsilon>0$ \cite{Lee2002,AbboudEtAl2018} or $O(t(k) n^c)$ for constant $c$ and  function $t$ of $k$ only \cite{DowneyAndFellows1999}. Our presented transformation passes these difficulties of $k$-Clique onto 
 problem $\beta$-retaining \mskt.

The algorithms for $\beta$-retaining \mskt\hs solving optimal Markov network approximation are also practically useful for solving other 
scientific problems. 
In particular, since Hamiltonian paths constitutes a special class of bounded-degree spanning trees, the algorithms can be tailored to solving problem {\it Hamiltonian path-retaining} MS$k$T. This is especially useful for revealing hidden higher-order relationships over a set of random variables that possess an indispensable total order relation, e.g., time series data, linguistic sentences, and biomolecule sequences. Indeed,  
our work has inspired a general framework for efficient and accurate prediction of biomolecule 3D structures. 
According to this framework, on a given molecule consisting of an ordered sequence of residues, an 
underlying probabilistic graph can be formulated in which random variables represent residues, edges for possible interactions between residues, and ``backbone edges'' connecting neighboring residues on the sequence. 
The objective of the 3D structure prediction problem is to find a spanning $3$-tree, which maximizes sum of mutual information of all involved $4$-cliques in the $3$-tree. Here the mutual information of four variables in a $4$-clique is defined based on its probabilistic mapping to a tetrahedron motif involving 4 residues in the biomolecule, where tetrahedrons are geometric building blocks of the 3D structure \cite{chang2022optimal}.

\par
\section{Preliminaries}

In this section, we introduce the notions of Markov networks, $k$-tree, tree-width, and Markov $k$-tree, which play essential roles in discussions throughout the paper.   

\subsection{Markov Networks} A finite system comprising $n$ random variables is denoted as $\bX = \{X_1, \ldots, X_n\}$, with the joint probability distribution function $P(\bX)$. While $P(\bX)$ may offer insights into dependence relationships among the variables, it is often not known due potentially high-dimensional relationships among the variables. The task we are interested in is to approximate the joint probability distribution $P(\bX)$ with $P_G(\bX)$, a joint distribution of $\bX$ in their binary relationships characterized by graph $G$. We call $G$ the {\it topology} of $P_G(\bX)$.

\begin{definition}\rm
Let $G = (\bX, E)$ be a non-directed probabilistic graph with vertex set $\bX$ consisting of random variables and edge set $E \subseteq \bX \times \bX$. 
$G$ is called a {\it Markov network} if, for the joint probability distribution $P_G(\bX)$, 
\begin{equation}
\label{markov}
(X_i, X_j) \notin E \Longrightarrow P_G(X_i, X_j | \bX\backslash\{X_i, X_j\} ) = P_G(X_i | \bX\backslash\{X_i, X_j\})P_G(X_j | \bX\backslash\{X_i, X_j\}) \quad 
\end{equation}
That is, variables $X_i$ and $X_j$ without sharing an edge are conditionally independent given the the rest of variables in the graph. 
\end{definition}
The property in (\ref{markov}) is called the {\it pairwise Markov property}. It is equivalent to the following presumably stronger condition, called  {\it global Markov property} as long as the distribution $P_G(\bX)$ are always positive\cite{Grimmett1973}. Specifically, for  subsets of variables $\bY, \bZ, \bS \subseteq \bX$ and 
separator $\bS$ of $\bY$ and $\bZ$ of graph $G$, 
\begin{equation}
P_G(\bY, \bZ | \bS) = P_G(\bY | \bS)P_G(\bZ | \bS) \quad 
\end{equation}

\subsection{$k$-tree and tree-width}

{\it Tree-width} is a metric on graphs that measures how much a graph is tree-like. There are a few alternative definitions for tree-width. The one most relevant to this work is via the notion of $k$-tree.

\begin{definition}
\label{k-tree-def}
\rm
Let $k \geq 1$ be an integer. The class of graphs called $k$-{\it trees} are defined recursively 
based on the number of vertices they contain:
\begin{itemize}
  \item[1.] A {\it $k$-tree} of exactly $k$ vertices  is a $k$-clique;
  
  \item[2.] A {\it $k$-tree} $G=(V, E)$ of $n$ vertices, for any $n>k$, consists of a {\it $k$-tree} $H=(U, F)$ of $n-1$ vertices and a vertex $v\not \in U$, such that, for some $k$-clique $C$ in $H$,
  
  (a) $V=U \cup \{v\}$; \\  
  (b) $E = F \cup \{ (v, u): u\in C\}$.
\end{itemize}
\end{definition}

When $k=1$, $k$-trees are 1-trees, which are simply trees under the regular sense \cite{BODLAENDER19981}.
However, for $k\geq 2$, $k$-trees are not trees as they contain cycles.
We call any subgraph of a $k$-tree a {\it partial $k$-tree}.



\begin{definition}\rm
Let $G$ be a graph. Then the {\it tree-width} of $G$ is the smallest number 
$k$ for which $G$ is a partial $k$-tree.
\end{definition}


Apparently any $k$-tree is a graph of tree-width $k$. The following assertion gives a stronger relationship between graphs of tree-width $k$ and $k$-trees.
\begin{proposition}
For every fixed $k\geq 1$, $k$-trees are maximum graphs of tree-width $k$.
\end{proposition}

As defined, a $k$-tree with $n$ vertices is generated through a recursive process that constructs a sequence of 
(sub)-$k$-trees whose numbers of vertices are $k, k+1, \dots, n$. During the generations, vertices and edges are created via rules 1 and 2 in Definition~\ref{k-tree-def}. This leads to the following further notes on the notion of $k$-tree.

First, by Definition~\ref{k-tree-def}, every vertex in a $k$-tree is generated with respect to a certain subset of other vertices. Typically, an application of rule 2 generates a new vertex and associates it with an existing clique of $k$ vertices to form a \kclq. For vertices in the $k$-clique generated by rule 1, no explicit association of a vertex to others is given. If vertices in the $k$-clique are considered generated one at a time, like by rule 2, then there is an assumed precedence order in the $k$-clique such that vertex $u$ precedes vertex $v$  if and only if $v$ is generated with respect to $u$.

\begin{definition}\rm
Let $G=(V, E)$ be a $k$-tree for some $k\geq 1$. A associated {\it precursor function} for $G$ is a mapping  $\pi: V \rightarrow 2^{V}$ that designates a subset of vertices for every vertex $v$, such that  

(1) if $v$ is generated with rule 2 of Definition~\ref{k-tree-def}, then $\pi(v) = C$, where $C$ is given by rule 2, or\\
\indd (2) if $v\in C_0$, the $k$-clique created with rule 1, then $\pi(v) \subseteq C_0\setminus \{v\}$; and $\forall u, v\in C_0$, $\pi(u) \subseteq \pi(v)$ \\
\indd\dind if and only if $\pi(v)\not \subseteq \pi(u)$.
\end{definition}

\vspace{-3mm}
Note that case (2) of precursor function $\pi$ imposes a total order for the vertices in set $C_0$ generated by rule 1 and hence there exists a vertex, named $v_1$, such that $\pi(v_1)=\emptyset$. Specifically, vertices in set $C_0$ are ordered as
$(v_1, v_2, \dots, v_{k})$ if and only if 
$\emptyset = \pi(v_1) \subset \pi(v_2) \subset  \dots \subset \pi(v_{k})$
where $v_i \in C_0$, $i=1,\dots, k$.

Second, different construction processes of the same $k$-tree may result in different associated {\it precursor function}. In this paper, when a $k$-tree is assumed, some associated precursor function  is also assumed.

Third, based on Definition~\ref{k-tree-def}, the total number of $(k+1)$-cliques in the $k$-tree of $n$ vertices, for $n>k$, is exactly $n-k$.

\subsection{Markov $k$-tree:}
\begin{definition}\rm 
Let $k \geq 1$ be an integer. A {\it Markov $k$-tree} is a Markov network over $n$ random variables $\bX = \{X_1, \ldots, X_n\}$ with a topology graph $G = (\bX, E)$ being a $k$-tree. The joint probability distribution function of the Markov $k$-tree is defined by $P_G(\bX)$ :

\begin{equation}
\label{markov-ktree}
   P_G(\bX) = \prod_{X_i \in \bX} P(X_i | \pi(X_i)) \quad
\end{equation}
\end{definition}
where $\pi$ is some associated precursor function\footnote{
Our previous  \cite{chang2022optimal} work proves that the joint probability defined for a given Markov $k$-tree is invariant of its associated precursor function $\pi$.}.  
Note that the joint probability $P_G(\bX)$ given in equation~\ref{markov-ktree} has the right-hand-side that is actually derived using the chain-rule of multivariate probability and conditional independence by Markov properties. Specifically, the right-hand-side is obtained by  following the reversed process of constructing the $k$-tree that corresponds to the associated precursor function $\pi$. 

\subsection{Mutual information}
Our discussions on approximation of Markov networks are  based on Shannon's information theory. 

\begin{definition}\rm
Let $\bX$ be a finite set of random variables with distribution $P(\bX)$. Then the {\it entropy} $H(\bX)$ of variables $\bX$ is defined as $H(\bX) = - \sum P(\bX) \log_2 P(\bX)$.
\end{definition}

While the Shannon's entropy accounts for the averaged number of (binary) bits to encode random variable values of one distribution, another notion of entropy can be used to measure the difference between two distributions \cite{KullbackAndLeibler1951}.
\begin{definition}\rm
Let $P$ and $Q$ be two distributions for the set $\bX$ of random variables. Then the {\it relative entropy} or {\it Kullback–Leibler divergence} between $P$ and $Q$ is defined as
\begin{equation}
\label{kl-divergence}
 D_{KL}(P\parallel Q) = \sum P(\bX) \log_2 \frac{P(\bX)}{Q(\bX)} 
 \end{equation}
\end{definition}

In particular, with $P(X_i, X_j)$ representing the joint probability between two random variables $X_i$ and $X_j$ and $Q(X_i, X_j) = P(X_i) P(X_j)$ representing their independence distribution, we obtain
\begin{definition}\rm
Let $P$ be a joint distribution for random variables $X_i$ and $X_j$. Their {\it mutual information} is defined as
\begin{equation}
\label{mi}
 I(X_i; X_j) = \sum P(X_i, X_j) \log_2 \frac{P(X_i, X_j)}{P(X_i)P(X_j)}
\end{equation}
\end{definition}
Equation~\ref{mi} can be extended to the following mutual information between a single variable $X_i$ and a set $\bY$ of random variables:
\begin{equation}
\label{mv-mi}
I(X_i; \bY) = \sum P(X_i, \bY) \log_2 \frac{P(X_i, \bY)}{P(X_i)P(\bY)}
\end{equation}

\section{Approximation with  $k$-tree topology}

We now discuss approximate of Markov networks with topologies specified with graphs. Let $\bX =\{X_1, \dots, X_n\}$ be a set of random variables and  $P(\bX)$ be a probabilistic
distribution of a Markov network over variables $\bX$
with an unknown topology. For graph $G=(V, E)$, with $|V|=n$, we denote with $P_G(\bX)$ the {\it distribution of a Markov network over random variables $\bX$ with topology} $G=(\bX, E)$.

\begin{definition}\rm 
Let integer $n\geq 1$ and ${\cal G}_n$ be a class of graphs (of $n$ vertices). 
The {\it optimal approximation of $P(\bX)$ with topology} ${\cal G}_n$ is a distribution $P_{G^*}(\bX)$ such that
\[ {G^*} = \arg \min_{G \,\in \,{\cal G}_n} D_{KL} (P(\bX) \parallel P_G(\bX))\]
\end{definition}

Chow and Liu \cite{chow1968approximating} initiated a seminal work on approximation with the topology class ${\cal T}_n$ of trees. They proved that for the tree class, the optimal approximation is via a tree topology that yields the maximum sum of mutual information $\sum_{X_i \in \bX} I(X_i; X_i')$, where $X_i'$ is the {\it predecessor} variable of variable
$X_i$. Such a tree can be found by the algorithm that solves the maximum spanning tree problem in linear time.

Our previous work \cite{chang2022optimal} extended the result to the topology class of $k$-trees and show that the optimal approximation can be achieved with a spanning $k$-tree topology yielding the maximum mutual information  
\begin{equation}
\label{max-k-tree}
\sum_{X_i \in \bX} I(X_i; \pi(X_i))
\end{equation}
for any associated precursor function $\pi$.

The derivation of the result in (\ref{max-k-tree}) is based on the connection between the minimization of the KL-divergence and the maximization of a spanning $k$-tree, as follows. Let $G$ be a spanning $k$-tree with an associated precursor function $\pi$. Applying equation~\ref{markov-ktree} to
the KL-divergence formula 
\begin{equation}
\label{KL-divergence}
D_{KL} (P(\bX) \parallel P_G(\bX)) = \sum P(\bX) \log_2 \frac{P(\bX)}{P_G(\bX)}
\end{equation}
 to yield 
\begin{equation}
\label{derivation-1}
\begin{split}
D_{KL} (P(\bX) \parallel P_G(\bX)) & =  \sum P(\bX) \log_2 \frac{P(\bX)}{P_G(\bX)} \\
& = \sum P(\bX) \Big ( \log_2  P(\bX) - \log_2 P_G(\bX) \Big) \\
& =  \sum P(\bX) \log_2  P(\bX) - \sum P(\bX) \sum_{X_i\in \bX}\log_2 P_G(X_i | \pi(X_i) \\
& = H(\bX) - \sum P(\bX) \sum_{X_i\in \bX} \log_2 P_G(X_i | \pi(X_i)
\end{split}
\end{equation}
where the $\sum P(\bX)$ is over all values of $\bX$ and the last term on the right-hand-side of the last equation in (\ref{derivation-1}) can be rewritten as
\begin{small}
\begin{equation}
\label{derivation-2}
\begin{split}
 &\sum \sum_{X_i \in \bX} P(\bX) \log_2 \frac{P_G (X_i, \pi(X_i)}
{P_G(\pi(X_i))}\\
 & = \sum_{X_i \in \bX} \sum P(\bX) \log_2 \frac{P_G (X_i, \pi(X_i)}{P_G(\pi(X_i))P_G(X_i)} + \sum_{X_i \in \bX} \sum P(\bX)\log_2 P_G(X_i)\\
 & =  \sum_{X_i \in \bX} \sum P(X_i, \pi(X_i)) \log_2 \frac{P_G (X_i, \pi(X_i)}{P_G(\pi(X_i))P_G(X_i)} + \sum_{X_i \in \bX} \sum P(X_i)\log_2 P_G(X_i)\\
& =  \sum_{X_i \in \bX} I(X_i; \pi(X_i)) - \sum_{X_i \in \bX} H(X_i)\\
\end{split}
\end{equation}
\end{small}

where the second last equality holds because $P(\bX)$ is projected on component $(X_i, \pi(X_i)$ and component $X_i$ with $P$ values over other components summed up to 1.

Because both $H(\bX)$ in (\ref{derivation-1}) and 
$\sum_{X_i\in \bX} H(X_i)$ in (\ref{derivation-2}) are invariant of the choice of topology $G$, combining equations~(\ref{derivation-2}) with (\ref{derivation-1}) leads to the conclusion that a topology $G$ to minimize distance $D_{KL} (P(\bX) \parallel P_G(\bX))$ if and only if it maximizes sum of mutual information $\sum_{X_i\in \bX} I(X_i; \pi(X_i))$, where $\pi$ is any associated precursor function with $G$. 

We now turn the latter problem into a graph-theoretic problem. Let $k\geq 1$ be an integer and $[n]$ be the positive integer set. 

\begin{definition}\rm
Let $f: [n]^{k+1} \rightarrow \mathbb{R}$ be a function. The problem {\it $f$-based maximum spanning $k$-tree}, denoted with \msktf, is defined as: given an underlying non-directed graph $G=(V, E)$, finds a maximum spanning $k$-tree $H$ of $G$ such that the objective function $\sum_{\Delta \in H} f(\Delta)$ achieves the maximum, where $\Delta \subseteq V$ is a $(k+1)$-clique in $H$. 
\end{definition}

Note that to make the problem \msktf\, well defined, the input graph $G$ needs to include values of function $f$ 
over some (but not necessarily all) $(k+1)$-cliques in the graph.  We often omit $f$ from the problem name \msktf, leaving it just \mskt, when $f$ is known in the context or without being specifically referred to in a discussion.

\begin{theorem}
The optimal approximation of Markov networks with $k$-tree topology can be achieved by solving the problem {\rm \msktf}, where for every $(k+1)$-clique $\Delta = \{X\}\cup \pi(X)$, $f(\Delta) = I(X; \pi(X))$.
\end{theorem}

Unfortunately, the intractability of the following problem makes \mskt\hs difficult to solve. 
\begin{proposition} [\cite{Leizhen1993}]
The following decision problem is NP-hard on every fixed $k\geq 2$: Determining if a given graph $G$ possesses a spanning $k$-tree as its subgraph.
\end{proposition}


\section{$\beta$-retaining MS$k$T}

We now turn into a restricted version of problem \mskt.  

\begin{definition}\rm
Let $\beta$ be a class of graphs. {\it $\beta$-retaining} \mskt\hs is the \mskt\hs problem whose outputted maximum spanning $k$-tree is required to
include some spanning subgraph $B\in \beta$ designated in the input graph $G$.
\end{definition}

By the definition, in the graph $G$ as the input for the {\it $\beta$-retaining} \mskt\hs problem, a spanning subgraph $B$ of $G$, which belongs to the class $\beta$, needs to be specified.  Our discussion will be focused on $\beta$ that is specifically the class of bounded-degree spanning trees. That is, problem {\it $\beta$-retaining \mskt} requires its outputted maximum spanning $k$-tree to retain the designated bounded-degree spanning tree from the input graph $G$.


\begin{lemma}
\label{edge}
Let $G$ be a $k$-tree with $n>k$ vertices. Then

{\rm (1)} Every edge in $G$ belongs to some \kclq\hs in $G$.\\
\indd {\rm (2)} During creation of $G$ with Definition~\ref{k-tree-def}, no edge $(u, v)$ can be created after $u$ and $v$  have \\
\dind \indd been created. That is, edge $(u, v)$ can only be created when either $u$ or $v$ being created.
\end{lemma}


\begin{definition}\rm
Let $G$ be a $k$-tree, $k\geq 1$, and $\Delta$ be any \kclq\hs in $G$. Then any \kclq\hs $\Delta'$ in $G$ is called a {\it neighbor of} $\Delta$ if $|\Delta \cap \Delta' | = k$.
\end{definition}

\begin{definition}\rm
Let a subset $S\subseteq V$  of vertices in graph $G=(V, E)$. $S$ {\it separates} the graph into two or more {\it disconnected components} if $S$ disjoins with all these components and 
any path connecting vertices between any pair of the disconnected components should contains some vertex in $S$.
\end{definition}

\begin{lemma}
\label{bounded-neighbors}
    Let $G=(V,E)$ be a $k$-tree and $\Delta$ be a $(k+1)$-clique in $G$. Assume that $\Delta$ separates  $G$ into $m$ connected components. Then the number of neighbors for $\Delta$ is at most $m$. 
\end{lemma}     
 
\noindent
{\bf Proof:}

    Let $\Delta$ be a $(k + 1)$-clique in the $k$-tree $G$. Denote with   
    ${\cal N}(\Delta)$ the set of all neighbors of $\Delta$. 

Let $\Delta_1, \Delta_2 \in {\cal N}(\Delta)$ be two neighbors of $\Delta$ and $z_1, z_2$ are two vertices such that $z_1 \in \Delta_1\backslash \Delta$ and $z_2 \in \Delta_2\backslash\Delta$. It is clear that $z_1 \not \in \Delta$ and $z_2 \not \in \Delta$. We claim that $z_1$ and $z_2$ cannot belong to the same connected component due to the separation of $G$ by the $(k+1)$-clique $\Delta$.
 
 To see the claim is true, consider otherwise there is a path $p=\{(z_1, y_1), (y_1, y_2), \dots, (y_{r}, y_{r+1})\}$, for some $r\geq 0$, connecting $z_1$ and $z_2$, where $y_{r+1}=z_2$, and the set $V(p)$ of vertices on $p$ is completely disjoint with $\Delta$. 
By Lemma~\ref{edge}, it is obvious that for any edge $(y_i, y_{i+1})$ in $p$, the edge has to be created along with either of its end vertices. This means either edge $(z_1, y_1)$ or $(y_{r}, z_2)$ is the last edge on $p$ to be created. 

\begin{figure}[h]
\begin{center}
\includegraphics[width=0.50\textwidth]{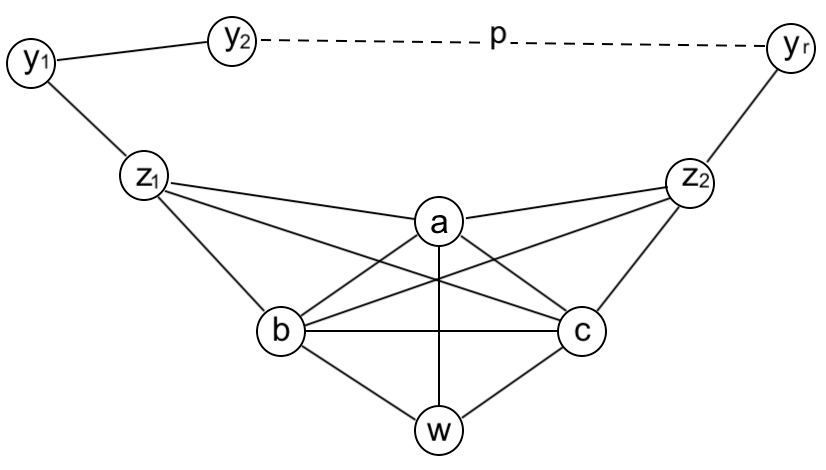}
\caption{A part of a 3-tree, containing three 4-cliques $\Delta=\{a,b, c, w\}$, 
$\Delta_1=\{a, b, c, z_1\}$, and $\Delta_2 =\{a, b, c, z_2\}$, to illustrate the argument in the proof of Lemma~\ref{bounded-neighbors} which claims any path disjoint with vertex set $\Delta$ but connecting $z_1$ and $z_2$ should not exist.  Path $p=\{(z_1, y_1), (y_1, y_2), \dots, (y_{r}, y_{r+1})\}$, $y_{r+1} = z_2$, is  assumed to be such a path; the assumption leads to a contradiction.}
\label{components}
\end{center}
\end{figure}

Without loss of generality, assume $(z_1, y_1)$ is the last edge on $p$ to be created. It implies that $z_2$ and $y_1$ have been created earlier. Then there are only two scenarios to consider. (1) \kclq\hs $\Delta_2$ was created along with $z_2$ being created. This implies that $k$-clique $\Delta_2 \cap \Delta_1$ had already existed and that consequently either 
\kclq\hs $\Delta_1$ or edge $(z_1, y_1)$ cannot be created. See Figure~\ref{components}. Contradicts. (2)  $z_2$ was created before  $(k+1)$-clique $\Delta_2$ is created. Based on the argument for case (1),  
$(k+1)$-clique $\Delta_2$ can only be created after $(k+1)$-clique $\Delta_1$, which also suggests that $\Delta_2$ cannot be created since $z_2$ has already existed. Contradicts.


The above argument shows that for any two vertices $z_1 \in \Delta_1$ and $z_2 \in \Delta_2$, where $\Delta_1, \Delta_2 \in {\cal N}(\Delta)$ are neighbors of \kclq\hs $\Delta$, $z_1$ and $z_2$ belong to two different connected components separated by $\Delta$. So if the number of connected components is $m$, 
$|{\cal N}(\Delta)| \leq m$. \epf


\begin{definition}\rm 
\label{bounded-branches}
A $k$-tree $G$ has {\it bounded branches}
 if every $(k + 1)$-clique $\Delta$ in $G$ has a bounded number of neighbors.
A graph $H$ is {\it bounded branching-friendly} if every $k$-tree that
contains $H$ as a spanning subgraph has bounded branches.
\end{definition}

\begin{theorem}
\label{beta}
Let $\beta$ be the class of bounded-degree trees. Then any graph in $\beta$ is bounded branching friendly.
\end{theorem}

\noindent
{\bf Proof}: Let $H\in \beta$ be any bounded-degree tree in $\beta$. We will show that every $k$-tree $G$ that contains $H$ as a spanning tree has bounded branches. That is to show that every \kclq\hs in $k$-tree $G$ separates $G$ into a bounded number of connected components, and then by Lemma~\ref{bounded-neighbors}, every \kclq\hs in such a $G$ has  bounded number of neighbors.

Assume constant $d\geq 1$ to be the highest degree of vertices in $H$ and let $\Delta$ be any \kclq\hs in the $k$-tree $G=(V, E)$ that contains $H$ as a spanning tree. Let $S\subseteq V$ be any subset of vertices, with $|S|=j$. We claim that the number of
connected components in $G$ separated by $S$ is at most $dj-j+1$. We prove the claim by induction on $j$.

When $j=1$, $S$ contains only one vertex, which can separate $H$ into at most $d$ components. Since $H$ is a spanning tree of $G$, the vertex can separate $G$ into at most $d=d\times 1 -1+1$ components
 
Assume $S$ contains $j=m$ vertices and it separates $G$ into at most $dm-m+1$ connected components. Consider now the case that $S$ contains $j=m+1$ vertices. Let vertex $v\in S$. By assumption, set $S\setminus \{v\}$ should separates $G$ into at most $dm-m+1$ connected components. Let $C$ be one such component that $v$ belongs to. Then $v$ can further separate subgraph $C\cap H$, thus subgraph 
$C\cap G$ into at most $d$ components. Together with the other $dm-m$ components, these are $dm-m+d = d(m+1) - (m+1) +1$ connected components separated by $S$ of $m+1$ vertices. 

Apply the proved claim to \kclq\hs $\Delta$, it is clear that $\Delta$ separates $G$ into at most $d(k+1)-k$
connected components. Since $\Delta$ is an arbitrary \kclq\hs in $G$, by Lemma~\ref{bounded-neighbors} and Definition~\ref{bounded-branches},   $G$ has bounded branches. And again by  Definition~\ref{bounded-branches}, we conclude that   $H$ in $\beta$ is bounded branching friendly.
 \epf

\begin{corollary} 
Let $\beta$ be the class of Hamiltonian paths. Then any graph in $\beta$ is bounded branching friendly.
\end{corollary}

\section{Algorithm}

In this section, we show that, for any bounded branching friendly class $\beta$ of graphs, the $\beta$-retaining MS$k$T problem can be solved in polynomial time for every fixed $k\geq 1$. We will discuss such an algorithm extensively for $\beta$ that is the class of bounded-degree spanning trees, which can be generalized to any bounded branching friendly class $\beta$.

\subsection{Tree-decomposition representation} 

We first introduce some further notations to facilitate discussions. 
Consider again Definition~\ref{k-tree-def} for $k$-trees. It is clear that every $k$-tree consists of a collection ${\cal K}$ of $(k+1)$-cliques. Every \kclq\hs is created as $C \cup \{v\}$, exclusively along with a newly introduced vertex $v$. On the other hand,  there may be more than one \kqq that contains the whole set $C$ of vertices. To establish succinct relationships between \kclq s, recursively we label every \kqq uniquely as $\Delta^{u}_{v}$, where $v$ is the newly introduced vertex for creation of the \kqq and $u\not=v$ satisfies: 

(1) either $u = \epsilon$, or \\
\indd (2) $\exists$ \kqq $\Delta^{x}_{u}$, for some $x \in V \cup \{\epsilon\}$ and $u\in V$, which is created earlier, with\\
\indd \dind   $|\Delta^{x}_{u} \cap \Delta^{u}_{v}| =k$.

In (2), $\Delta^{x}_{u}$ is called the {\it parent} of $\Delta^{u}_{v}$, denoted with $\rho(\Delta^{u}_{v}) = \Delta^{x}_{u}$. It is clear that every \kq, except the one first created (that does not have a parent), has a unique parent.

\begin{proposition}
The relation defined by {parent} $\rho$ between \kq s in $k$-tree $G$ forms a rooted-tree topology $({\cal K}_G, {\cal T}_G)$, where

(1) ${\cal K}_G=\{\Delta: \Delta \mbox{ is a \kclq\hs in $G$ }\}$;\\
\indd (2) $\Delta^{\epsilon}_x \in {\cal K}_G$, for some $x\in V$, is the designated {\it root};\\
\indd (3) ${\cal T}_G = \{(\Delta^x_u, \Delta^u_v): \Delta^x_u, \Delta^u_v \in {\cal K}_G, 
\,
\Delta^x_u = \rho(\Delta^u_v)\}$
\end{proposition}

We call $({\cal K}_G, {\cal T}_G)$ a {\it tree-decomposition}\footnote{The notion of tree-decomposition was first introduced independently from the notion of $k$-tree \cite{RobertsonAndSeymour1983}. It is used in this section solely for the purpose of discussion our algorithms on $k$-trees.} of $k$-tree $G$. We have the following important property for tree-decomposition.

\begin{proposition}
\label{td-property}
Let $\dl'$ and $\dl''$ be two \kclq s on a {\it tree-decomposition} of $k$-tree $G$. Let vertex $v\in \dl' \cap \dl''$. Then $v \in \dl$ for every \kqq $\dl$ on the path between $\dl'$ and $\dl''$ on the tree.
\end{proposition}
\noindent
{\bf Proof}: Since $\dl$ is on the path between $\dl'$ and $\dl''$ on the tree, there are only two scenarios about their creations. (1)  $\dl$ is an ``ancestor'' of both $\dl'$ and $\dl''$. (2) $\dl$ is a descendent of one and an ancestor of another. In both scenarios, if $v\not \in \dl$, there are duplicated creations of vertex $v$, contradicting that $k$-tree definition. 
\epf

The notation of \td offers a higher level view on a $k$-tree and makes the discussion easier on algorithms for $k$-tree optimization. 
In particular, construction of a $k$-tree is equivalent to construction of the \td of the corresponding $k$-tree.

Specifically, the \td of a $k$-tree is a tree rooted at \kqq $\Delta^{\epsilon}_x$ for some $x\in V$, with tree nodes, both internal and leaf nodes, drawn from set ${\cal K}_G$. Since the tree is rooted, it is directional. Therefore, 
\kqq $\dl^u_v$ is an internal node with a child node $\dl^v_y$, for any $y\in V$, if and only if $\rho(\dl^v_y) = \dl^u_v$.  Formally, we need the following technical results in the next section.

\begin{definition}
\rm
Let $({\cal K}_G, {\cal T}_G)$ be a {\it tree-decomposition} of $k$-tree $G=(V,E)$, with root $\dl^{\epsilon}_x$, $x\in V$. For any $\dl \in {\cal K}_G$, we define 
subset $V_{\dl}$ of $V$, such that 
\[ V_{\dl} = \{ v: v \in \dl', \dl' \in {\cal K}_G, \mbox{ $\dl'$ is a descendent of $\dl$ in $({\cal K}_G, {\cal T}_G)$}\}\]
\end{definition}

Clearly, if $\dl$ is the root node of any tree decomposition, then $V_{\dl} = V$.
 \begin{proposition}
 \label{sub-td}
 Let $({\cal K}_G, {\cal T}_G)$ be a {tree-decomposition} of $k$-tree $G=(V,E)$. Then 
 the subtree rooted at \kqq $\dl$ is a {tree-decomposition} of the induced subgraph of the $k$-tree $G$ by vertex set $V_\dl$.
 \end{proposition}

 

\subsection{Dynamic programming}

By Proposition~\ref{sub-td}, a \td for $k$-tree $G$ can be constructed by building sub-tree-decompositions and assemble them into the whole tree-decomposition. This leads to the following repetitive (thus recursive) process to construct a \td. Specifically, let $\dl^u_v$ be an internal node of the tree with children 
$\dl^v_{w_i}$, $i=1,2\dots, m$, for some $m\geq 1$. Building the subtree rooted at $\dl^u_v$ with 
the children can be done by first connecting child $\dl^v_{w_m}$ to $\dl^u_v$ and then continuing to build 
the subtree rooted at $\dl^u_v$ with the rest of the children $\dl^v_{w_i}$, $i=1,2\dots, m-1$.

We now discuss the details. To create the child $\dl^v_{w_i}$ from parent node $\dl^u_v$, vertex $w_i$ in $G$ can only be one that was not created before. This means that the creation process needs to keep the record of an exact set from which a new vertex can be drawn for creation of a subtree rooted at $\dl^u_v$. Because $\beta$ is a class of bounded branches friendly graphs, by the proofs of Lemma~\ref{bounded-neighbors} and of Proposition~\ref{td-property}, this set of vertices is actually partitioned into at most $c$ exclusive subsets, for some constant $c\geq 1$; vertices from only one of these subsets can be drawn for creation of the subtree. In other words, these subsets correspond to some of the connected components of $H$ separated by the \kqq $\dl^u_v$. Since $H$ is a known subgraph given in the input graph $G$, vertices in every one of these subsets are known when $\dl^u_v$ is specified. This suggests that every subset can be retrieved with an index identifier and there are a constant number of such indexes for these exclusive subsets. 

Specifically, assume $\lambda \geq 1$ to be the largest number such that $H$ is separated 
by any \kqq into at most $\lambda$ connected components. Let $[\lambda]$ denote the set $\{1,2,\dots, \lambda]$. Given tree node (i.e., \kclq) $\dl^u_v$, let $S: [\lambda] \rightarrow 2^V$ denote  
the function that maps any identifier $i\in [\lambda]$ to the set of vertices in the $i^{\rm th}$ connected component (resulted from the separation by $\dl^u_v$). Then the set of vertices in the subtree rooted at $\dl^u_v$, excluding those in $\dl^u_v$, can be written as  $V_{u,v, \cali} = \bigcup_{i\in \cali} S(i)$, for some subset $\cali \subseteq [\lambda]$. We will show that that set $V_{u,v,\cali}$ is uniquely determined by $\cali$ along with $\dl^u_v$ during the tree construction process.

\begin{definition}
\rm
\label{function-m}
Let $f$ be a real-value function with the argument being  a \kclq. We define real-value function $F(\dl^u_v, \cali)$ to be the maximum sum of $f$ values over all \kclq s ({\it excluding} $\dl^u_v$)  in a subtree rooted at $\dl^u_v$ that contains vertices in 
$V_{u,v, \cali} \cup \dl^u_v$.
\end{definition}

Then we can derive the following recurrence for function $F$: 
\begin{equation}
\label{dp}
 F(\dl^u_v, \cali) = \max_{i\in \cali,\, w\in S(i)\setminus \dl^u_v} \Big\{ F(\dl^v_w, {\cal J}_i) + F(\dl^u_v, \cali\setminus\{i\}) + f(\dl^v_w) \Big\}
\end{equation}

where ${\cal J}_i\subseteq [\lambda]$ is a subset of identifiers for the connected components of $H$ separated by $\dl^v_w$.

The recurrence~(\ref{dp}) has the base cases $F(\dl^u_v, \emptyset) = 0$, for all $u, v\in V$.

\begin{lemma}
\label{dp-correctness}
Recurrence~(\ref{dp}) together with its base cases computes correctly function $F(\dl^u_v, \cali)$.
\end{lemma}
\noindent
{\bf Proof}: We prove the claim by showing that the recurrence computes correctly the sum of 
$f$ scores on all \kclq s  in the subtree rooted at $\dl^u_v$, excluding $\dl^u_v$ itself. 
We do this by induction on $h$, the number of \kclq s in the subtree rooted at $\dl^u_v$.

When $h=1$, the subtree contains only  \kqq $\dl^u_v$ itself. This corresponds to the case case 
$F(\dl^u_v, \emptyset)$ which has value 0. 

We hypothesize that when $h\leq N$, the claim is correct. Consider the case $h=N+1$. Note that node $\dl^u_v$, the root of the subtree, 
has $|\cali|$ children nodes, each of which is the root of the subtree containing exactly the vertices in one of the 
connected components separated by $\dl^u_v$. 
The recurrence first adds term $f(\dl^v_w)$ into the sum, where $\dl^v_w$ is one of the children node and the root of the subtree covering the $i^{\rm th}$ connected component. Term $F(\dl^v_w, {\cal J}_i)$ recursively computes 
the sum of $f$ scores on the rest of \kclq s in the subtree covering the $i^{\rm th}$ connected component. Term  
$F(\dl^u_v, \cali\setminus\{i\})$ recursively computes the sum of $f$ scores on the rest of subtrees rooted at  
$\dl^u_v$. In both cases, the numbers of \kclq s to compute are at most $N$ and they are both correct by the hypothesis.

Now we justify that given \kqq $\dl^u_v$ and identifiers $\cali$, the information of vertices belonging to  connected components (of $H$ separated by $\dl^u_v$) can be determined. Consider vertices in $\dl^u_v$ are identified on $H$,  upon which a search process (e.g., depth-first search) on $H$ can determine connected components. Vertices are associated with these components and each component is associated with an identifier in $\cali \subseteq [\lambda]$. The component that vertex $u$ belongs, which can be determined, should be in set $\cali$. The same argument applies to ${\cal J}_i$.   
\epf

Since Recurrence~(\ref{dp}) maximizes the sum of $f$ scores on all \kclq s of the subtree  rooted at a \kqq except the root, the answer to the  $\beta$-retaining \mskt$(f)$ problem can be expressed as:
\begin{equation}
\label{mskt-solution}
 \arg \max_{\dl^{\epsilon}_u \in V^{k+1}, \, \cali \subseteq [\lambda]} \Big\{F(\dl^{\epsilon}_u, \cali) + f(\dl^{\epsilon}_u)\Big\}
 \end{equation}
by examining all possible roots $\dl^{\epsilon}_u$ for a maximum spanning $k$-tree.

Function $F$ can be implemented with a dynamic programming algorithm based on the recurrence  equation~(\ref{dp}). This essentially involves establishing a look-up table that computes values of function $F$ on all necessary combinations of $(\dl^u_v, \cali)$. For graph of $n$ vertices, the table size is $O(n^{k+1} \times 2^\lambda)$ and each entry in the table can be computed in time $O(n)$ for selecting vertex $w\in S(i)$. This gives rise to the time complexity 
$O(n^{k+2})$.

\begin{theorem}
\label{algo-mskt}
Let $\beta$ be the class of all bounded branches friendly graphs. The $\beta$-retaining \mskt$(f)$ problem can be solved in time $O(n^{k+2})$, for every fixed $k\geq 1$.
\end{theorem}

\begin{theorem}
Given any Markov network of $n$ random variables, its $k$-tree topology approximation with the minimum loss of information can be computed in time $O(n^{k+2})$, for every fixed $k\geq 1$, provided that the found topology retains some designated spanning graph that is bounded branches friendly.
\end{theorem}

\begin{corollary}
There are polynomial time algorithms for the minimum loss of information approximation of Markov networks with 
a $k$-tree topology that retains a  designated Hamiltonian path in the input network.
\end{corollary}

\begin{corollary}
There are polynomial time algorithms for the minimum loss of information approximation of Markov networks with 
a $k$-tree topology that retains a designated bounded-degree spanning tree in the input network.
\end{corollary}

\subsection{Optimality in complexity}

An early study \cite{DingThesis2016} on the  $\beta$-retaining \mskt\, problem showed that, for 
the class $\beta$ of Hamiltonian paths, the algorithm can be tweaked such that  
the time complexity can be improved to $O(n^{k+1})$. 
We believe the technique and the time bound can be generalized to all bounded branches friendly
classes $\beta$. We omit such a proof which is rather lengthy.

However, the polynomial time $O(n^{k+1})$ for $\beta$-retaining \mskt\, may not be further improved in term of its exponent parameter $k$. In the following we show strong evidence that this claim is true.
For this we defined the following decision problem related to $\beta$-retaining \mskt$(f)$ with $\beta$ being the class of Hamiltonian paths. We assume $f$ is real-value function with argument being a \kclq\hs together with weights of edges they share. 

\hmsktf:\\
\dind \ind Input: edge-weighted graph $G$ with an Hamiltonian path $H$, integer $k\geq 1$,  real number $S$;\\
\dind \ind Output: ``Yes'' if and only if there is a spanning $k$-tree retaining $H$ whose sum of $f$ scores\\
\dind \dind \dind \dind  over all \kclq s is at least $S$;

We now connect problem \hmsktf\hs to the classical graph-theoretic problem $k$-Clique, which determines if the input graph has a clique of size as the given threshold $k$. While $k$-Clique problem can be solved trivially in time  $O(n^k)$ on graph of $n$ vertices, any substantial improvement to the upper bound has proved extremely difficult. In particular, it is unlikely \cite{Lee2002,AbboudEtAl2018} to solve problem $k$-Clique in time $O(n^{k-\epsilon})$ for any $\epsilon>0$. In addition, it has also been proved  \cite{DowneyAndFellows1999} that solving $k$-Clique in time $t(k)n^c$, for some function $t(k)$ independent of $n$ and some constant $c$ would lead to an unlikely breakthrough in computational complexity theory. We will show such difficulties for $k$-Clique also passes on to 
 \hmsktf\hs and thus to the problem $\beta$-retaining \msktf\hs for all bounded branches friendly classes $\beta$. This complexity connection from $k$-Clique is through a transformation from $k$-Clique to \hmsktf, which is a ``parameterized reduction'' in the sense that the parameter $k$ in $k$-Clique is directly translated to the parameter $k$ in \hmsktf, independent of the input graph size $n$. 

\begin{proposition}
\label{reduction}
Problem $k$-Clique can be transformed to problem \hmsktf\hs via a parameterized reduction.
\end{proposition}
\noindent
{\bf Proof}: We sketch the desired reduction. Given an input instance for $k$-Clique: a non-directed graph $G=(V, E)$ and a parameter $k$,  an instance for \hmsktf\hs is constructed, which includes an edge-weighted  graph $G'$, Hamiltonian path $H$ in $G'$, parameter $k'$, and score threshold $\sigma$. Specifically, 

(1) Graph $G' = (V', E')$, where $V'=\{l(x_i): x_i \in V\}$, where $l: V \rightarrow [n]$, where $n=|V|$,  $l$ is an \\\dind \dind one-to-one mapping, and $E'=V'\times V'$. I.e., graph $G'$ is a complete, vertex-labeled graph. 

\vspace{-2mm}
(2) Edge weights are defined by function $w: E' \rightarrow  \mathbb{R}$, such that for every edge $(l(u,), l(v))\in E'$, 
\[w(l(u), l(v)) = \begin{cases} 
1 & (u, v) \in E\\
0 & (u, v) \not \in E\\
\end{cases}
\]

\vspace{-4mm}

(3) The designed Hamiltonian path in $G'$ is $H=\{(i, i+1): (i, i+1) \in E', 1\leq i < n\}$; 
\vspace{-2mm}

(4) Parameter $k' = k-1$;
\vspace{-2mm}

(5) Score threshold $\sigma=1$.

Our proof choose real-value function $f$ such that for any $(k'+1)$-clique $\dl$, $f(\dl)= \prod\limits_{x,y\in \dl} w(x,y)$.

Now if $G$ has a clique of size $k$, let it be $\{x_1, x_2, \dots, x_k\}$. Then the corresponding set of vertices in $G$, a $(k'+1)$-clique $\dl= \{l(x_1), l(x_2), \dots, l(x_k)\}$, has score $f(\dl) = 1$. Since $G'$ is complete graph, there is a spanning $k'$-tree formed by $\dl$ and additional $n-k'$ $(k'+1)$-cliques, which
includes all edges in $H$ and has the sum of $f$ scores at least $\sigma=1$. On the other hand, if $G$ has a spanning $k'$-tree containing $H$ with the sum of $f$ scores at least at $S$, one of the $(k'+1)$-cliques, 
say $\{l(x_1), l(x_2), \dots, l(x_k)\}$,
in the $k'$-tree should have all edges with weight 1. This is translated to that $\{x_1, x_2, \dots, x_k\}$ is a clique in original graph $G$, by the edge weight definition for $G'$. Thus $G$ has a clique of size $k$. \epf

Proposition~\ref{reduction} shows that our algorithm demonstrated earlier is likely to be the most efficient for the $\beta$-retaining \mskt$(f)$ problem and for optimal approximation of Markov networks with $k$-tree topology.

\subsection{Conclusions}

We have demonstrated polynomial-time algorithms to solve the $\beta$-retaining maximum spanning $k$-tree problem, for every fixed integer $k\geq 1$. Our research has revealed that a maximum spanning $k$-tree topology corresponds to an optimal approximation of Markov networks (with the minimum information loss). While finding a maximum spanning $k$-tree is computationally intractable, we have shown that the problem can be solve efficiently when the desired $k$-tree also retains a designated spanning subgraph in 
graph classes of certain characteristics, namely being bounded branches friendly. We also demonstrated strong evidence that our algorithms are likely the optimal in time complexity.
 
In this paper, we have considered two classes $\beta$ of graphs: bounded-degree spanning trees and its subclass of Hamiltonian paths. The \mskt\hs problems that retains graphs from these two classes have arisen from practical applications, for example, a biomolecule 3D structure graph containing the backbone as a designated Hamiltonian path \cite{chang2022optimal} and a gene or metabolic network containing a known, critical pathway as a designated spanning tree \cite{ChatterjeeAndPal2016}. As the maximum spanning $k$-tree problem is essential to network approximation, it is of interest to investigate larger classes of graphs that can be retained in finding maximum spanning $k$-tree. One such class may be 2-connected graphs \cite{Schrijver2003} that lie between the class of trees and and class of $2$-trees. 
\clearpage

\bibliographystyle{IEEEtran}
\bibliography{Approx-MSkT.bib}

\begin{thebibliography}{10}
\providecommand{\url}[1]{#1}
\csname url@samestyle\endcsname
\providecommand{\newblock}{\relax}
\providecommand{\bibinfo}[2]{#2}
\providecommand{\BIBentrySTDinterwordspacing}{\spaceskip=0pt\relax}
\providecommand{\BIBentryALTinterwordstretchfactor}{4}
\providecommand{\BIBentryALTinterwordspacing}{\spaceskip=\fontdimen2\font plus
\BIBentryALTinterwordstretchfactor\fontdimen3\font minus
  \fontdimen4\font\relax}
\providecommand{\BIBforeignlanguage}[2]{{%
\expandafter\ifx\csname l@#1\endcsname\relax
\typeout{** WARNING: IEEEtran.bst: No hyphenation pattern has been}%
\typeout{** loaded for the language `#1'. Using the pattern for}%
\typeout{** the default language instead.}%
\else
\language=\csname l@#1\endcsname
\fi
#2}}
\providecommand{\BIBdecl}{\relax}
\BIBdecl

\bibitem{KschischangEtAl2001}
F.~R. Kschischang, B.~J. Frey, and H.~A. Loeliger, ``Factor graphs and the
  sum-product algorithm,'' \emph{IEEE Transactions on Information Theory},
  vol.~47, no.~2, pp. 498--519, 2001.

\bibitem{BleiEtAl2017}
D.~M. Blei, A.~Kucukelbir, and J.~D. McAuliffe, ``Variational inference: A
  review for statisticians,'' \emph{Journal of the American Statistical
  Association}, vol. 112, no. 518, pp. 859--877, 2017.

\bibitem{Murphy2012}
K.~Murphy, \emph{Machine learning: a probabilistic perspective}.\hskip 1em plus
  0.5em minus 0.4em\relax MIT Press, 2012.

\bibitem{FosterAndKesselman2003}
I.~Foster and C.~Kesselman, \emph{The Grid: Blueprint for a New Computing
  Infrastructure}.\hskip 1em plus 0.5em minus 0.4em\relax Morgan Kaufmann,
  2003.

\bibitem{GionisEtAl1999}
A.~Gionis, P.~Indyk, and R.~Motwani, ``Similarity search in high dimensions via
  hashing,'' in \emph{Proceedings of Conference on Vvery Large Databases},
  1999, pp. 518---529.

\bibitem{mohseni2014rolling}
T.~Mohseni~Ahooyi, J.~E. Arbogast, and M.~Soroush, ``Rolling pin method:
  efficient general method of joint probability modeling,'' \emph{Industrial \&
  Engineering Chemistry Research}, vol.~53, no.~52, pp. 20\,191--20\,203, 2014.

\bibitem{MELUCCI201991}
M.~Melucci, ``A brief survey on probability distribution approximation,''
  \emph{Computer Science Review}, vol.~33, pp. 91--97, 2019.

\bibitem{7869051}
J.~Jiao, Y.~Han, and T.~Weissman, ``Beyond maximum likelihood: Boosting the
  chow-liu algorithm for large alphabets,'' in \emph{The 50th Asilomar
  Conference on Signals, Systems and Computers}, 2016, pp. 321--325.

\bibitem{chitsaz2009birna}
H.~Chitsaz, R.~Backofen, and S.~C. Sahinalp, ``bi{RNA}: Fast {RNA}-{RNA}
  binding sites prediction,'' in \emph{Proceedings of 9th International
  Workshop in Algorithms in Bioinformatics}.\hskip 1em plus 0.5em minus
  0.4em\relax Springer, 2009, pp. 25--36.

\bibitem{chow1968approximating}
C.~Chow and C.~Liu, ``Approximating discrete probability distributions with
  dependence trees,'' \emph{IEEE transactions on Information Theory}, vol.~14,
  no.~3, pp. 462--467, 1968.

\bibitem{kovacs2009approximation}
E.~Kov{\'a}cs and T.~Sz{\'a}ntai, ``On the approximation of a discrete
  multivariate probability distribution using the new concept of t-cherry
  junction tree,'' in \emph{Coping with Uncertainty: Robust Solutions}.\hskip
  1em plus 0.5em minus 0.4em\relax Springer, 2009, pp. 39--56.

\bibitem{gene}
J.~Suzuki, ``A novel {Chow--Liu} algorithm and its application to gene
  differential analysis,'' \emph{International Journal of Approximate
  Reasoning}, vol.~80, pp. 1--18, 2017.

\bibitem{montiel2013approximating}
L.~V. Montiel and J.~E. Bickel, ``Approximating joint probability distributions
  given partial information,'' \emph{Decision Analysis}, vol.~10, no.~1, pp.
  26--41, 2013.

\bibitem{Speech}
N.~Hammami and M.~Bedda, ``Improved tree model for {Arabic} speech
  recognition,'' in \emph{The 3rd International Conference on Computer Science
  and Information Technology}, vol.~5, 2010, pp. 521--526.

\bibitem{1202220}
K.~Huang, I.~King, and M.~Lyu, ``Constructing a large node {Chow-Liu} tree
  based on frequent itemsets,'' in \emph{Proceedings of the 9th International
  Conference on Neural Information Processing,}, vol.~1, 2002, pp. 498--502
  vol.1.

\bibitem{king2009mist}
B.~M. King and B.~Tidor, ``Mist: Maximum information spanning trees for
  dimension reduction of biological data sets,'' \emph{Bioinformatics},
  vol.~25, no.~9, pp. 1165--1172, 2009.

\bibitem{KullbackAndLeibler1951}
S.~Kullback and R.~Leibler, ``On information and sufficiency,'' \emph{Annals of
  Mathematical Statistics}, vol.~22, no.~1, pp. 79--86, 1951.

\bibitem{9719784}
E.~Boix-Adser{\`a}, G.~Bresler, and F.~Koehler, ``{Chow-Liu}{++}: Optimal
  prediction-centric learning of tree ising models,'' in \emph{IEEE 62nd Annual
  Symposium on Foundations of Computer Science}, 2022, pp. 417--426.

\bibitem{tan2010learning}
V.~Y. Tan, S.~Sanghavi, J.~W. Fisher, and A.~S. Willsky, ``Learning graphical
  models for hypothesis testing and classification,'' \emph{IEEE Transactions
  on Signal Processing}, vol.~58, no.~11, pp. 5481--5495, 2010.

\bibitem{zhang2019tree}
X.~Zhang, D.~Chang, W.~Qi, and Z.~Zhan, ``Tree-like dimensionality reduction
  for cancer-informatics,'' in \emph{IOP Conference Series: Materials Science
  and Engineering}, vol. 490, no.~4.\hskip 1em plus 0.5em minus 0.4em\relax IOP
  Publishing, 2019, p. 042028.

\bibitem{jiao2016beyond}
J.~Jiao, Y.~Han, and T.~Weissman, ``Beyond maximum likelihood: Boosting the
  chow-liu algorithm for large alphabets,'' in \emph{2016 50th Asilomar
  Conference on Signals, Systems and Computers}.\hskip 1em plus 0.5em minus
  0.4em\relax IEEE, 2016, pp. 321--325.

\bibitem{clustering}
C.~Chan and T.~Liu, ``Clustering by multivariate mutual information under
  chow-liu tree approximation,'' in \emph{2015 53rd Annual Allerton Conference
  on Communication, Control, and Computing (Allerton)}, 2015, pp. 993--999.

\bibitem{STEIMER2015520}
A.~Steimer, F.~Zubler, and K.~Schindler, ``{Chow--Liu} trees are sufficient
  predictive models for reproducing key features of functional networks of
  periictal eeg time-series,'' \emph{NeuroImage}, vol. 118, pp. 520--537, 2015.

\bibitem{wainwright2008graphical}
M.~J. Wainwright, M.~I. Jordan \emph{et~al.}, ``Graphical models, exponential
  families, and variational inference,'' \emph{Foundations and Trends in
  Machine Learning}, vol.~1, no. 1--2, pp. 1--305, 2008.

\bibitem{karger2001learning}
D.~R. Karger and N.~Srebro, ``Learning markov networks: maximum bounded
  tree-width graphs.'' in \emph{SODA}, vol.~1, 2001, pp. 392--401.

\bibitem{bresler2015efficiently}
G.~Bresler, ``Efficiently learning ising models on arbitrary graphs,'' in
  \emph{Proceedings of the forty-seventh annual ACM symposium on Theory of
  computing}, 2015, pp. 771--782.

\bibitem{singh2015finding}
A.~Singh and M.~D. Humphries, ``Finding communities in sparse networks,''
  \emph{Nature Scientific Reports}, vol.~5, no.~1, p. 8828, 2015.

\bibitem{gunduz2004cell}
C.~Gunduz, B.~Yener, and S.~H. Gultekin, ``The cell graphs of cancer,''
  \emph{Bioinformatics}, vol.~20, no. suppl\_1, pp. i145--i151, 2004.

\bibitem{humphries2008network}
M.~D. Humphries and K.~Gurney, ``Network `small-world-ness': a quantitative
  method for determining canonical network equivalence,'' \emph{PloS one},
  vol.~3, no.~4, p. e0002051, 2008.

\bibitem{newman}
M.~E. Newman, ``The structure and function of complex networks,'' \emph{SIAM
  review}, vol.~45, no.~2, pp. 167--256, 2003.

\bibitem{Patil1986}
H.~Patil, ``On the structure of k-trees,'' \emph{Journal of Combinatorics,
  Information and System Sciences}, vol.~11, no. 2-4, pp. 57--64, 1986.

\bibitem{RobertsonAndSeymour1983}
N.~Robertson and P.~D. Seymour, ``Graph minors. i. excluding a forest,''
  \emph{Journal of Combinatorial Theory, Series B}, vol.~35, no.~1, pp. 39--61,
  1983.

\bibitem{BODLAENDER19981}
H.~L. Bodlaender, ``A partial k-arboretum of graphs with bounded treewidth,''
  \emph{Theoretical Computer Science}, vol. 209, no.~1, pp. 1--45, 1998.

\bibitem{Srebro2003}
N.~Srebro, ``Maximum likelihood bounded tree-width markov networks,''
  \emph{Artificial Intelligence}, vol. 143, pp. 123--138, 2003.

\bibitem{SzantaiAndKovacs2012}
T.~Sz\'{a}ntai and K.~E., ``Hypergraphs as a mean of discovering the dependence
  structure of a discrete multivariate probability distribution,'' \emph{Annals
  of Operations Research}, vol. 193, pp. 71--90, 2012.

\bibitem{chang2022optimal}
D.~Chang, L.~Ding, R.~Malmberg, D.~Robinson, M.~Wicker, H.~Yan, A.~Martinez,
  and L.~Cai, ``Optimal learning of markov k-tree topology,'' \emph{Journal of
  Computational Mathematics and Data Science}, vol.~4, p. 100046, 2022.

\bibitem{Leizhen1993}
L.~Cai and F.~Maffray, ``On the sprnning k-tree problem,'' \emph{Discrete
  Applied Mathematics}, vol.~44, no. 1-3, pp. 139--156, 1993.

\bibitem{Lee2002}
L.~Lee, ``Fast context-free grammar parsing requires fast boolean matrix
  multiplication,'' \emph{Journal of ACM}, vol.~49, no.~1, pp. 1--15, 2002.

\bibitem{AbboudEtAl2018}
A.~Abboud, A.~Backurs, and V.~V. Williams, ``If the current clique algorithms
  are optimal, so is valiant's parser,'' \emph{SIAM Journal of Computing},
  vol.~47, no.~6, 2018.

\bibitem{DowneyAndFellows1999}
R.~Downey and M.~Fellows, \emph{Parameterized Complexity}.\hskip 1em plus 0.5em
  minus 0.4em\relax Springer, 1999.

\bibitem{Grimmett1973}
G.~R. Grimmett, ``A theorem about random fields,'' \emph{Bulletin of the London
  Mathematical Society}, vol.~5, no.~1, pp. 81--84, 1973.

\bibitem{DingThesis2016}
L.~Ding, ``Maximum spanning k-trees: models and applications,'' PhD
  Dissertation, University of Georgia, 2016.

\bibitem{ChatterjeeAndPal2016}
P.~Chatterjee and N.~R. Pal, ``Discovery of synergistic genetic network: A
  minimum spanning tree-based approach,'' \emph{Journal of Bioinformatics and
  Computational Biology}, vol.~14, no.~1, 2016.

\bibitem{Schrijver2003}
A.~Schrijver, \emph{Combinatorial Optimization}.\hskip 1em plus 0.5em minus
  0.4em\relax Springer, 2003.

\end{thebibliography}

\end{document}